\documentclass[11pt]{article}
\usepackage{latexsym}  % for Symbol fonts
\usepackage{amssymb}
\usepackage{graphicx}
\usepackage{amsmath}
\usepackage{mathrsfs}
\newcommand{\overarc}[1]{\stackrel{\Large\mbox{$\frown$}}{#1}}
\newcommand{\be}{\begin{equation}}
\newcommand{\ee}{\end{equation}}
\newcommand{\bea}{\begin{eqnarray}}
\newcommand{\eea}{\end{eqnarray}}
\newcommand{\bref}[1]{(\ref{#1})}

\begin{document}
\begin{titlepage}
%%%%% PREPRINT NUMBERS %%%%%%
%\begin{flushright}
%\today
%\end{flushright}
%%%%%%%%%%%%%%%%%%%%%%%%%%%%%%
%\vspace{3\baselineskip}
%PACS numbers: 13.15.+g, 12.10.-g, 14.60.-z

%%%%%%%%%%%%%%%%%%% TITLE %%%%%%%%%%%%%%%%%%
\begin{center}
{\Large\bf  
The Relativistic Corrections of GPS	
}
\end{center}

\begin{center}

%%%%%%%%%%%%%%%% AUTHORS %%%%%%%%%%%%%%%%%%%%%%%
\vspace{0.1cm}

%%%%%%%%%%%%%%%%
{\large Takeshi Fukuyama$^{a,}$%
\footnote{E-mail: fukuyama@rcnp.osaka-u.ac.jp}},
%{\large Nobuchika Okada$^{b,}$%
%\footnote{E-mail: okadan@ua.edu}}
and
{\large Sueo Sugimoto$^{b,}$%
\footnote{E-mail: sugimoto@se.ritsumei.ac.jp}}

%%%%%%%%%%%%%%%%%%%%%%% AFFILIATION %%%%%%%%%%%%
\vspace{0.2cm}

${}^{a} ${\small \it Research Center for Nuclear Physics (RCNP),
Osaka University, \\Ibaraki, Osaka, 567-0047, Japan}\\

${}^{b}${\small \it Department of Electrical Engineering, Ritsumeika University,  Kusatu, Shiga, 525-8577, Japan}

%\medskip
%\vskip 10mm

\end{center}

\begin{abstract}
We calculate the special and general relativistic effects of the Global Positioning System (GPS), especially the effects depending on the small deviation
of the orbit from the circular one. This effect is well known but, to our knowledge,  the detail of its derivation is not known publicly. 
GPS is widely used in particle physics experiments and this letter may be useful among this community.

\end{abstract}
%PACS number: 12.10.-g, 12.10.Dm, 12.10.Kt
\end{titlepage}
\section{Introduction}
The transmitter (Space Vehcle or Satellite) and receiver suffer many kinds of time corrections. 
Indeed, the time delays of Space Vehcle (SV) are summarized as \cite{200K}
\be
\Delta t_{SV} = a_{f0} + a_{f1}(t-t_{oc}) + a_{f2}(t-t_{oc})^2 + \Delta t_r
\ee
See page $92$ of \cite{200K} for the definitions of $a_{fi}$ and $t_{oc}$. These are irrelevant to this paper. Relevant is the last term $\Delta t_r$, which is the the special and general relativistic effects due to the small deviation of the real elliptic orbit from the complete circular motion with the same energy. The relativistic time delay in the latter case is orbit-position independent and took into consideration beforehand but that in the former case is position dependent and must be calculated at each position of SV (hereafter we call it satellite). This deviation is well known as
\cite{200K}
\be
\Delta t_r=-2\frac{\sqrt{GM_\oplus a}}{c^2}(E-M)=-2\frac{\sqrt{GM_\oplus a}}{c^2}e\sin E.
\label{PEQ}
\ee
Here, $G, M_\oplus, a, e, c, E, M$ are gravitational constant, mass of the earth, the major semi-axis of the elliptic orbit, its eccentricity, light velocity, eccentric anomaly, mean anomaly, respectively, whose detailed explanation is given later in Fig.1.
The purpose of this letter is to derive this equation explicily.
\section{Special relativistic effect}
In the inertial frames, the infinitesimal world interval of light propagation is given by
\be
ds^2=c^2dt^2-dx^2-dy^2-dz^2=0
\label{SR1}
\ee
in any inertial frames with $c=3\times 10^8$ m/s. This interval is also described as
\bea
ds^2&=&(cdt,~dx,~dy,~dz)
\left(
\begin{array}{cccc}
1&0&0&0\\
0&-1&0&0\\
0&0&-1&0\\
0&0&0&-1
\end{array}
\right)\left(
\begin{array}{c}
cdt\\
dx\\
dy\\
dz
\end{array}
\right)\nonumber\\
&\equiv & \sum_{\mu,\nu=0}^{3}\eta_{\mu\nu}dx^\mu dx^\nu
\label{SR2}
\eea
with $x^0=ct,~x^1=x,~x^2=y,~x^3=z$.
In the following, we will discuss the case where the potential energy depends only on the distance $r$ from a fixed point.
In this case it is more comvenient to adopt the spherical coordinates. In the spherical coordinates, Eq.\bref{SR2} is represented as
\be
ds^2=c^2dt^2-dr^2-r^2(d\theta^2+\sin^2\theta d\phi^2)=g_{\mu\nu}dx^\mu dx^\nu,
\label{SR3}
\ee
where
\be
g_{\mu\nu}=\left(
\begin{array}{cccc}
1&0&0&0\\
0&-1&0&0\\
0&0&-1&0\\
0&0&0&-\sin^2\theta
\end{array}
\right)
\ee
and $x^1=r,~x^2=r\theta,~x^3=r\phi$. Let us consider a train passing a platform with constant velocity ${\bf v}$.  Consider the world distance of a fixed point $p$ in the train from two frames, one fixed with the platform (denoting quantities without dash) and the other fixed with the train (with dash). The invariance of the world intervals gives
 \be
c^2dt'^2=c^2dt^2-dx^2-dy^2-dz^2\equiv c^2dt^2-dl^2=\left(1-\frac{v^2}{c^2}\right)c^2dt^2.
\ee
That is, the clock $t'$ in the train runs slow relative to $t$ in the platform by
\be
dt'=\sqrt{1-\frac{v^2}{c^2}}dt\approx \left(1-\frac{v^2}{2c^2}\right)dt.
\ee
Thus the clock moving with velocity $v$ runs slow by $\left(1-\frac{v^2}{2c^2}\right)$ per second relative to that in the rest frame,
\be
\frac{\delta t}{t}=-\frac{v^2}{2c^2}.
\ee
\section{General relativistic effect}
In this section, we consider the effect of gravitation. Let us consider two bodies motion with the earth with mass $M_\oplus$ and the satellite with mass mass $m (\ll M_\oplus)$. The time delay in the satellite is considered in general relativity with weak field approximation.
In this case, as we know from text books on general relativity \cite{Fuku1}, we obtain the following relativistic corrections,
\bea
ds^2&=&(1-\frac{r_g}{r})c^2dt^2-(1+\frac{r_g}{r})dr^2-r^2(d\theta^2+\sin ^2\theta d\phi^2)\nonumber\\
&\approx &\left(1-\frac{r_g}{r}-\frac{v^2}{c^2}\right)c^2dt^2=c^2dt'^2.
\eea
Here $r_g\equiv 2GM_\oplus/c^2$. $t'$ is the clock fixed at the satellite, where both velocity and gravitation are zeros. The satellite takes a free falling motion and there the gravitational force vanishes. Therefore, the clock in the satellite runs slow by
\be
\frac{\delta t}{t}=-\frac{1}{2}\left(\frac{v^2}{c^2}+\frac{r_g}{r}\right).
\label{delay}
\ee
\section{The deviation of the orbit from the circular motion}
If the satellite orbit is exactly spherical, $r$ and $v$ are constants.
However, the orbit is generally elliptical and they are not constants. The force between the satellite and the earth is dominated by the central force
\be
{\bf F}=-\frac{GM_\oplus m}{r^2}\hat{{\bf r}}, ~~\hat{\bf r}\equiv \frac{{\bf r}}{r}.
\ee
In this system the angular momentum ${\bf L}\equiv {\bf r}\times {\bf p}$ is conserved, together with the satellite total dynamical energy,
\be
H=\frac{1}{2}mv^2-\frac{GM_{\oplus}m}{r},
\ee
where we denote the energy by $H$, distinguishing it from the eccentric anomaly $E$.

. That is, the motion is coplanner and the sectorial velocity is conserved. This remains true even in the general relativity. However, in real elliptical orbits, neither $r$ nor $v$ are constants.  If the relativistic corrections are performed in every stage of the orbit it is OK even in this case.  In the real GPS, the time delay due to circular motion is set on initially and the deviation from circular motion is taken into account by measuring the eccentric angle $E$ from the perigee (See Fig.1).  
\begin{figure}[h]
\centering
\includegraphics[scale=0.4]{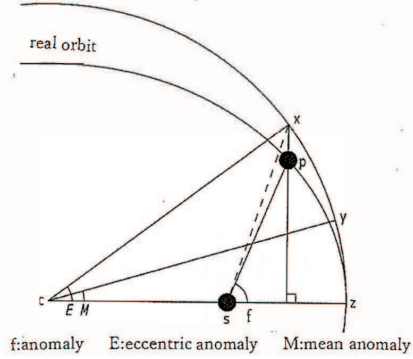}
\caption{The deviation of the real elliptic orbit from the circular one.}
\label{fig:Elliptic}
\end{figure}
Hereafter we denote the time delay due to the deviation from the circular motion by $\Delta$. Then
\be
\frac{\Delta t}{t}=-\frac{1}{2}\left(\frac{2{\bf v}\cdot\Delta{\bf v}}{c^2}+r_g\Delta\frac{1}{r}\right).
\ee
Let us consider that the satellite is at the position $p$ at $t$ after it passed the perigee $z$ at $t_z$. (See Fig.1).
The real orbit is the ellipse whose measure and minor semi-axses are $a$ and $b$, respectively,
\be
a=\frac{p}{1-e^2}=\frac{GM_\oplus m}{2|H|},~~b=\frac{p}{\sqrt{1-e^2}}=\frac{L}{\sqrt{2m|H|}}.
\label{orbit}
\ee
Here $L\equiv |{\bf L}|$ and $p$ is the semiparameter, $p\equiv \frac{L^2}{m^2GM_\oplus}$. It should be remarked that major semi-axis depends only on the energy $H$ and not on $L$. In Fig.1, the mean anomaly $M$ is 
\be
M\equiv \frac{2\pi}{T}(t-t_z),
\ee
where $T$ is the orbit period, $T=\pi GM_\oplus m\sqrt{\frac{m}{2|H|^3}}$. (Hereafter we set $t-t_z$ as $t$.)  Therefore, the point $y$ indicates the position if it took the circular motion. The eccentric anomaly $E$ (therefore, the position $x$) is defined by Fig.1. This $E$ is proved to satisfy the Kepler's law,
\be
E-M=e\sin E,
\label{Kepler}
\ee
by using the integrals $H$,  ${\bf L}$ and the additional vector integral,
\be
{\bf v}\times{\bf L}-GM_\oplus m\frac{{\bf r}}{r}=\text{const vector}.
\label{integral}
\ee
The last integral is the peculiar one appearing in the special form of potential, $U(r)\propto 1/r$ \cite{LL} among generally $r^n$ potentials. Using equations of motion, you can easily check that the time derivative of Eq.\bref{integral} indeed vanishes.
The $E$ satisfies the nonlinear equation Eq.\bref{Kepler} and is meassured from the final result of the recurrence formula
\be 
E_{j+1}=E_j-\frac{E_j-e\sin E_j- M}{e\cos E_j-1}.
\label{Kepler2}
\ee

From the fact that  the circle in Fig.1 is tangential to the ellipse at $z$ and from Eq.\bref{orbit}, the real elliptical and circular orbits have the same energy $H$,
\be
H_{ellipse}=\frac{1}{2}mv_{ellipse}(t)^2-\frac{GM_\oplus m}{r_{ellipse}(t)}=H_{circle}=\frac{1}{2}mv_{circle}^2-\frac{GM_\oplus m}{r_{circle}}\equiv H.
\label{equal}
\ee
Therefore $\Delta H=0$, Namely,
\be
\Delta\left(\frac{1}{2}mv^2-\frac{GM_\oplus m}{r}\right)=\frac{mc^2}{2}\Delta\left(\frac{v^2}{c^2}-\frac{r_g}{r}\right)=0.
\label{equal2}
\ee
Then the deviation effects of velocity $v$ and $r$ are same. First, we calculate the deviation effect of velocity.
\be
\frac{\Delta t_v}{t}=-\frac{{\bf v}\cdot\Delta {\bf v}}{c^2}=-\frac{v}{c^2}\frac{\Delta r}{t}=-\frac{1}{c^2}v\frac{a(E-M)}{t}.
\label{Deltav}
\ee
The second equality comes from that fact that $\Delta v\equiv \frac{\overarc{xy}}{t}$ by definition. 

Also the velocity is approximated as the mean velocity 
\be
\overline{v}=\sqrt{\frac{GM_\oplus}{a}}.
\label{av}
\ee
Together with Eq.\bref{Kepler} and Eq.\bref{av}, Eq.\bref{Deltav} gives
\be
\Delta t_v=-\frac{1}{c^2}\sqrt{GM_\oplus a}e\sin E.
\label{200K1}
\ee
The calculation of the deviation effect of $\frac{-1}{2}\Delta \left(\frac{r_g}{r}\right)(\equiv \Delta t_p)$ is not so simple as that of $\frac{-1}{2}\Delta\left(\frac{v^2}{c^2}\right)(\equiv \Delta t_v)$.
Fortunately, we know from Eq.\bref{equal2} that $\Delta t_p=\Delta t_v$.
Thus we obtain Eq.\bref{PEQ},
\be
\Delta t_v+\Delta t_p=2\Delta t_v=-2\frac{1}{c^2}\sqrt{GM_\oplus a}e\sin E.
\ee

The precision of the formula Eq.\bref{PEQ} depends on the final difference $E_{j+1}-E_j$ in Eq.\bref{Kepler2}. It depends on the algorithm not concerned with the relativistic effects \cite{Numerical}.

\end{document}